\documentclass[12pt]{article}
\usepackage{bm}
\usepackage[usenames,dvipsnames]{xcolor}
\usepackage{srcltx}
\usepackage{graphicx}
\usepackage{epstopdf}
\usepackage{color}
\usepackage{amsmath}
\usepackage{amssymb}
\usepackage{color}
\usepackage{float}
\usepackage{subfigure}
\usepackage{gensymb}

\usepackage{epstopdf}
\usepackage{epsfig}
\usepackage{psfrag}
\usepackage[mode=errorstop]{pstool}
    \newcommand{\figref}{Fig.~\ref}
  	
 	\newcommand{\secref}{Sec.~\ref}

\begin{document}

\title{Metasurface Magnetless Specular Isolator}

\author{Guillaume Lavigne, Toshiro Kodera and Christophe Caloz}

\maketitle

\begin{abstract}
We present a (nongyrotropic) metasurface magnetless specular isolator. This device reflects as a mirror a wave incident under a specified angle in one direction and absorbs it in the opposite direction. The metasurface is synthesized in terms of bianisotropic susceptibility tensors, whose nonreciprocity resides in normal components and exhibits a hybrid electric, magneto-electric nature.
The metaparticle is implemented in the form of a U-shaped conducting structure loaded by a transistor. The operation principle of the specular isolator is demonstrated by both full-wave simulation and experiment, with isolation levels reaching $41$ and $38$~dB respectively. This system represents the first realization of a metasurface involving nonreciprocal normal susceptibilities and features a previously unreported type of nonreciprocity.
\end{abstract}

\section*{Introduction}\label{sec:intro}

Magnetless nonreciprocity has recently arisen as a solution for breaking Lorentz reciprocity without the drawbacks of the dominant magnetized ferrite or terbium gallium garnet~(TGGs) technologies~\cite{lax1962microwave,villaverde1978terbium,reiskarimian2021review}, namely incompatibility with integrated circuits, large size, heavy weigth and high cost. Magnetless nonreciprocity can be achieved in linear or nonlinear forms. The latter, being restricted to fixed intensity ranges and non-simultaneous excitations in opposite directions~\cite{shi2015limitations,sounas2018broadband}, does not represent a viable solution for practical devices~\cite{caloz2018electromagnetic}. In contrast, linear nonreciprocity  may be highly efficient, while bearing potential for novel types of nonreciprocities. In the microwave and millimeter-wave regimes, it subdivides into space-time modulated systems~(dynamic bias)~\cite{hadad2016breaking,shi2017optical,sounas2013giant,sounas2018angular,wang2018time,shaltout2015time,taravati2020full} and transistor-loaded systems~(static bias)~\cite{popa2007architecture,yuan2009zero,popa2012nonreciprocal,kodera2011artificial,wang2012gyrotropic,sounas2012electromagnetic,kodera2013magnetless,ra2016magnet,taravati2017nonreciprocal,kodera2018unidirectional,ra2020nonreciprocal,lavigne2021magnetless,taravati2021full,li2022nonreciprocal}. The transistor approach is particularly suitable for typical, monochromatic nonreciprocal operations (isolation, circulation and nonreciprocal phase-shifting), given their simple, low consumption and inexpensive (DC)~biasing scheme, and immunity to spurious harmonics and intermodulation products.

Metasurfaces, which have already led to myriad of novel electromagnetic applications~\cite{glybovski2016metasurfaces,achouri2018design}, represent an  unprecedented opportunity for magnetless nonreciprocity. This opportunity largely stems from the great diversity associated with bianisotropic metasurface designs, which provides superior control over the fundamental properties of electromagnetic waves~\cite{pfeiffer2014bianisotropic,epstein2016arbitrary,asadchy2016perfect,lavigne2018susceptibility,chen2018theory,asadchy2018bianisotropic}, and from the recent development of corresponding powerful synthesis techniques~\cite{achouri2015general,epstein2016huygens,achouri2020electromagnetic}. A number of magnetless nonreciprocal metasurface have been recently reported in transistor-loaded technology, including metasurfaces realizing nonreciprocal polarization rotation in reflection~\cite{kodera2011artificial,wang2012gyrotropic,lavigne2021magnetless} and in transmission~\cite{ra2016magnet,li2022nonreciprocal}, reflective spatial circulation~\cite{ra2020nonreciprocal}, transmissive isolation~\cite{taravati2017nonreciprocal} and nonreciprocal reflective beamsteering~\cite{taravati2021full}. However, these metasurfaces are either purely theoretical~\cite{ra2016magnet,ra2020nonreciprocal,lavigne2021magnetless}, or limited in terms of functionality~\cite{kodera2011artificial,wang2012gyrotropic} or yet relying on antenna-array technology~
\cite{taravati2017nonreciprocal,taravati2021full,li2022nonreciprocal}.


This paper reports a transistor-loaded magnetless nonreciprocal metasurface providing specular isolation, i.e. reflecting the wave incident from one direction as a mirror and absorbing the wave incident from the opposite direction. The related asymmetric reflection coefficient is realized by leveraging nonreciprocal normal metasurface susceptibilities~\cite{achouri2019angular,abdolali2019parallel}. The metasurface is synthesized using Generalized Sheet Transition Conditions (GSTCs) and a corresponding transistor-loaded metaparticle is proposed. The specular isolation operation is demonstrated by both full-wave simulation and prototype measurement.

\section*{Results}
\subsection*{Specular Isolator Concept}

\begin{figure}[h!]
  \includegraphics[width=\linewidth]{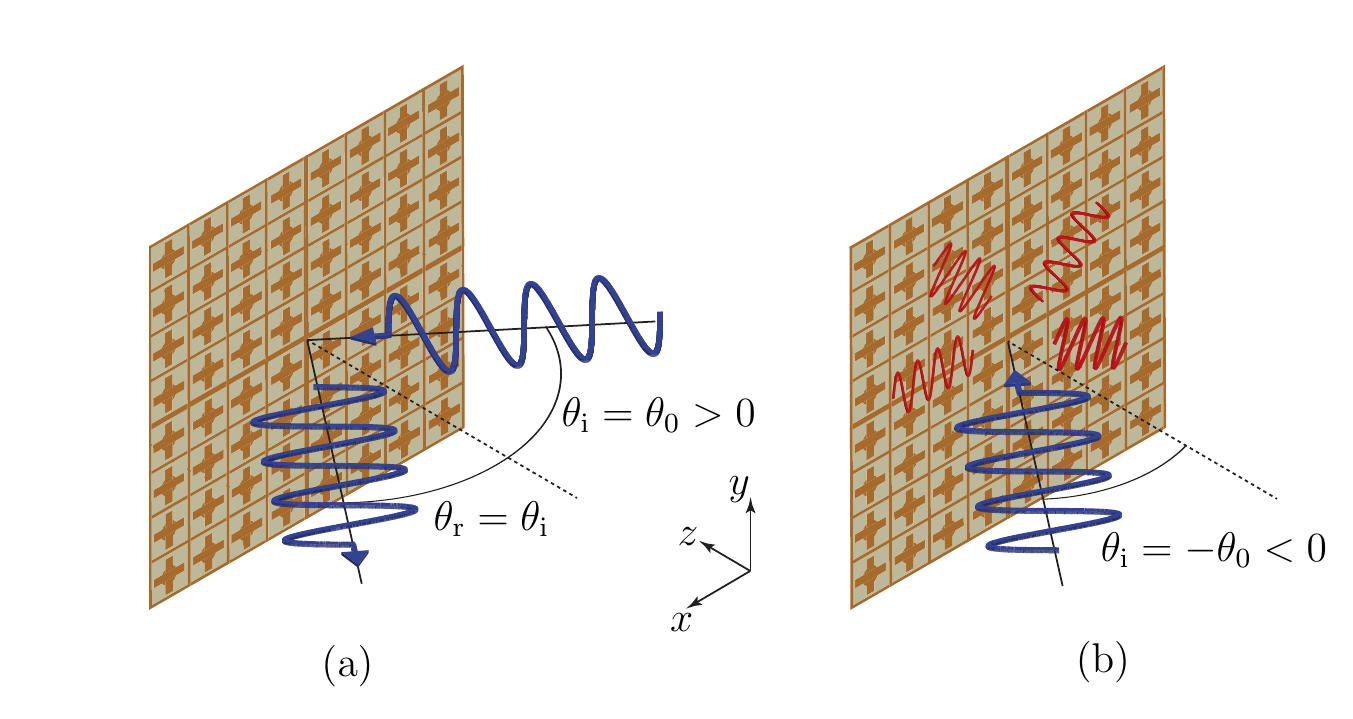}
  \caption{Concept of the metasurface specular isolator. (a) A wave incident at an angle $\theta_\text{i} = \theta_0$ is specularly reflected. (b) A wave incident at an angle $\theta_\text{i} =-\theta_0$ is absorbed by the structure without reflections.}\label{fig:concept}
\end{figure}

Figure~\ref{fig:concept} depicts the concept of the proposed metasurface specular isolator. A wave incident in the $xz$-plane at an angle $\theta_\text{i}= \sin^{-1}(k_{x\text{,i}}/ k)= \theta_0$, with $k = \omega/c$, where $\theta_0$ is the operation angle, is specularly reflected, while a wave incident at the angle $\theta_\text{i}=-\theta_0$ is absorbed by the metasurface. 

This operation implies the following three conditions on the metasurface. First, the specular nature of the reflection ($\theta_\text{r} = \theta_\text{i}$) requires that the metasurface has no phase gradient, which implies that it must be uniform. Second, the fact that there is only one scattered wave, and hence no diffraction orders, implies, assuming a periodic metasurface structure of period $d$, that $d <  |\lambda/ (\sin \theta_\text{i} \pm 1)|$. Third, the spatial asymmetry of the reflection implies breaking Lorentz reciprocity, which means that the metasurface must be nonreciprocal.

In addition, we assume that the metasurface is nongyrotropic, i.e., that it does not rotate the polarization of the incident wave upon reflection. Moreover, we shall consider only the p-polarized problem with incidence in the $xz$-plane, whose nonzero electromagnetic field components are $E_x$, $E_z$ and $H_y$, the s-polarized problem being solvable in an analog fashion.

\subsection*{Required Susceptibility Components}\label{sec:required_susceptibility}
Given its deeply subwavelength thickness, a metasurface can be conveniently modeled by the so-called Generalized Sheet Transition Conditions (GSTCs)~\cite{idemen2011discontinuities,kuester2003averaged}. The GSTCs read
  \begin{subequations}\label{eq:GSTCs_normal}
  \begin{equation}
    \hat{z} \times \Delta \mathbf{H} = j \omega \mathbf{P}_{\|} - \hat{z} \times \nabla M_z,
\end{equation}
\begin{equation}
    \hat{z} \times \Delta \mathbf{E} = -j \omega \mathbf{M}_{\|} - \frac{1}{\epsilon} \hat{z} \times \nabla P_z,
    \end{equation}
  \end{subequations}
  where the $\Delta$ symbol represents the difference of the fields on both sides of the metasurface, and where $\mathbf{P}$ and $\mathbf{M}$ are the induced electric and magnetic surface polarization densities, which may be expressed as
  \begin{subequations}\label{eq:polarizabilities_to_susceptibilities}
  \begin{equation}
   \mathbf{P} = \epsilon \overline{\overline{\chi}}_\text{ee} \mathbf{E}_\text{av} + \frac{1}{c} \overline{\overline{\chi}}_\text{em} \mathbf{H}_\text{av},
\end{equation}
\begin{equation}
   \mathbf{M} = \mu \overline{\overline{\chi}}_\text{mm} \mathbf{H}_\text{av} + \frac{1}{c} \overline{\overline{\chi}}_\text{me} \mathbf{E}_\text{av},
    \end{equation}
  \end{subequations}
where the ``av" subscript represents the difference of the field on both sides of the metasurface and $\overline{\overline{\chi}}_\text{ee}$, $\overline{\overline{\chi}}_\text{em}$, $\overline{\overline{\chi}}_\text{me}$ and $\overline{\overline{\chi}}_\text{mm}$ are the bianisotropic surface susceptibility tensors characterizing the metasurface~\cite{achouri2015general,achouri2020electromagnetic}. Assuming that the metasurface is placed in the $xy$-plane at $z=0$, the field differences and averages are
\begin{equation}\label{eq:av_diff}
\Phi_\text{av} =  \frac{\Phi_{\text{0}+} +\Phi_{\text{0}-}}{2}, \quad \Delta \Phi = \Phi_{\text{0}+} -\Phi_{\text{0}-},
\end{equation}
where $\Phi$ denotes either the electric or magnetic field.

In general, each of the susceptibility tensors in~\eqref{eq:polarizabilities_to_susceptibilities} includes $3 \times 3 = 9$ components, which leads to a total of $4 \times 9 = 36$ independent parameters. In the case of the proposed specular isolator, these components should be independent on $x$ and $y$ to satisfy the metasurface uniformity condition. Moreover, the required nonreciprocity implies the global condition~\cite{achouri2020electromagnetic,caloz2018electromagnetic}
\begin{equation}\label{eq:reciprocity_susceptibility}
  \overline{\overline{\chi}}_\text{ee} \neq \overline{\overline{\chi}}_\text{ee}^T, \quad \overline{\overline{\chi}}_\text{mm} \neq\overline{\overline{\chi}}_\text{mm}^T \quad \text{or} \quad \overline{\overline{\chi}}_\text{em} \neq - \overline{\overline{\chi}}_\text{me}^T.
\end{equation}
Finally, the nongyrotropy and p-polarized incidence assumptions eliminate 24 out of the 36 susceptibility components, simplifying the bianisotropic susceptibility tensors to
\begin{subequations}\label{eq:chi_tensor}
\begin{equation}
\overline{\overline{\chi}}_\text{ee} = \begin{bmatrix}
                                         \chi_\text{ee}^{xx} & 0 & \chi_\text{ee}^{xz} \\
                                         0 & 0 & 0 \\
                                         \chi_\text{ee}^{zx} & 0 & \chi_\text{ee}^{zz}
                                       \end{bmatrix}, \quad
\overline{\overline{\chi}}_\text{em} = \begin{bmatrix}
                                         0 & \chi_\text{em}^{xy} &  0\\
                                         0 & 0 & 0 \\
                                         0 & \chi_\text{em}^{zy} & 0
                                       \end{bmatrix},
\end{equation}
\begin{equation}
\overline{\overline{\chi}}_\text{me} = \begin{bmatrix}
                                         0 & 0 & 0 \\
                                         \chi_\text{me}^{yx} & 0 & \chi_\text{me}^{yz} \\
                                         0 & 0 & 0
                                       \end{bmatrix}, \quad
\overline{\overline{\chi}}_\text{mm} = \begin{bmatrix}
                                         0 & 0 &  0\\
                                         0 & \chi_\text{mm}^{yy} & 0 \\
                                         0 & 0 & 0
                                       \end{bmatrix},
\end{equation}
\end{subequations}
which include overall 9 parameters, where the nonreciprocity condition~\eqref{eq:reciprocity_susceptibility} translate into
\begin{equation}\label{eq:scalar_NR_relations}
  \chi_\text{ee}^{xz} \neq \chi_\text{ee}^{zx}, \quad \chi_\text{em}^{xy} \neq -\chi_\text{me}^{yx} \quad \text{or} \quad \chi_\text{em}^{zy} \neq -\chi_\text{me}^{yz}.
\end{equation}
The corresponding metasurface for the s-polarization would involve the dual susceptibility components $\chi_\text{mm}^{xx}$, $\chi_\text{mm}^{xz}$, $\chi_\text{mm}^{zx}$, $\chi_\text{mm}^{zz}$, $\chi_\text{em}^{yx}$, $\chi_\text{em}^{yz}$, $\chi_\text{me}^{xy}$, $\chi_\text{me}^{zy}$ and $\chi_\text{ee}^{yy}$, with the nonreciprocity condition $\chi_\text{mm}^{xz} \neq \chi_\text{mm}^{zx}$, $\chi_\text{em}^{yx} \neq -\chi_\text{me}^{xy}$ or $\chi_\text{em}^{yz} \neq -\chi_\text{me}^{zy}$ .

Inserting~\eqref{eq:polarizabilities_to_susceptibilities} into~\eqref{eq:GSTCs_normal} with~\eqref{eq:chi_tensor} leads the following explicit scalar GSTC equations
\begin{subequations}\label{eq:GSTC_simplified}
  \begin{equation}
    \Delta H_y = -j \omega \epsilon \chi_\text{ee}^{xx} E_{\text{av},x} - j \omega \epsilon \chi_\text{ee}^{xz} E_{\text{av},z} - j k \chi_\text{em}^{xy} H_{\text{av},y},
  \end{equation}
  \begin{equation}
  \begin{split}
    \Delta E_x = &-j \omega \mu \chi_\text{mm}^{yy} H_{\text{av},y} -j k \chi_\text{me}^{yx} E_{\text{av},x} -j k \chi_\text{me}^{yz} E_{\text{av},z} \\ & -\chi_\text{ee}^{xz} \partial_x E_{\text{av},z}- \chi_\text{ee}^{zz} \partial_x E_{\text{av},z}- \eta \chi_\text{em}^{zy} \partial_x H_{\text{av},y},
    \end{split}
  \end{equation}
\end{subequations}
where $\partial_x $ denotes the spatial derivative versus $x$. Assuming plane wave incidence, which allows the substitution $\partial_x \rightarrow -j k_x$, where $k_x = k \sin \theta$, reduces then~\eqref{eq:GSTC_simplified} to
\begin{subequations}\label{eq:GSTC_simplified_kx}
  \begin{equation}
    \Delta H_y = -j \omega \epsilon \chi_\text{ee}^{xx} E_{\text{av},x} - j \omega \epsilon \chi_\text{ee}^{xz} E_{\text{av},z} - j k \chi_\text{em}^{xy} H_{\text{av},y},
  \end{equation}
  \begin{equation}
  \begin{split}
    \Delta E_x =& -j \omega \mu \chi_\text{mm}^{yy} H_{\text{av},y} -j k \chi_\text{me}^{yx} E_{\text{av},x} -j k \chi_\text{me}^{yz} E_{\text{av},z} \\ &+ j k_x \chi_\text{ee}^{xz} E_{\text{av},z} + j k_x \chi_\text{ee}^{zz} E_{\text{av},z} + j k_x \eta \chi_\text{em}^{zy} H_{\text{av},y},
    \end{split}
  \end{equation}
\end{subequations}
which are the final GSTC equations for the problem at hand.

In these relations, the field differences and averages are found from~\eqref{eq:av_diff} in terms of the reflection and transmission coefficients, $R$ and $T$. Assuming incidence in the $+z$ direction, theses quantities read
\begin{subequations}\label{eq:av_diff_R_T}
  \begin{equation}
    \Delta E_x = \frac{k_z}{k}(T-1-R),
  \end{equation}
  \begin{equation}
    \Delta H_y =(-1+R+T)/\eta,
  \end{equation}
  \begin{equation}
    E_{\text{av},x} =\frac{k_z}{2 k}(1+R+T),
  \end{equation}
  \begin{equation}
   E_{\text{av},z} =\frac{k_x}{2 k}(1+T-R),
  \end{equation}
  \begin{equation}
   H_{\text{av},y} =(1+T-R)/2\eta,
  \end{equation}
\end{subequations}
 where $k_z = k \cos \theta$. Substituting~\eqref{eq:av_diff_R_T} into~\eqref{eq:GSTC_simplified_kx}, and solving for $R$ gives~\cite{achouri2019angular}
\begin{subequations}\label{eq:reflection_coefficient}
\begin{equation}
\begin{split}
  R = & \frac{2}{C}\{k_x^2 \chi_\text{ee}^{zz}-k_z^2 \chi_\text{ee}^{xx} -k_z[k_x (\chi_\text{ee}^{xz}-\chi_\text{ee}^{zx})- k (\chi_\text{em}^{xy}-\chi_\text{me}^{yx})] \\
  & -k k_x (\chi_\text{em}^{zy}+\chi_\text{me}^{yz}) + k^2 \chi_\text{mm}^{yy}\},
\end{split}
\end{equation}
where
\begin{equation}\label{C}
\begin{split}
  C = & 2 [ k_z^2 \chi_\text{ee}^{xx} + k_x^2 \chi_\text{ee}^{zz} - k k_x (\chi_\text{em}^{zy} + \chi_\text{me}^{yz})+k^2 \chi_\text{mm}^{yy}] \\
  &+ k^2(\chi_\text{ee}^{xx} \chi_\text{mm}^{yy} - \chi_\text{em}^{xy} \chi_\text{me}^{yx}) - j k_z \{k_x^2 ( \chi_\text{ee}^{xz} \chi_\text{ee}^{zx} - \chi_\text{ee}^{xx} \chi_\text{ee}^{zz}) \\
  &+4 - k k_x [\chi_\text{ee}^{zx} \chi_\text{em}^{xy} + \chi_\text{ee}^{xz} \chi_\text{me}^{yx} - \chi_\text{ee}^{xx}(\chi_\text{em}^{zy}+\chi_\text{me}^{yz})]\}.
\end{split}
\end{equation}
\end{subequations}

Realizing the specular isolation operation (see~\figref{fig:concept}) requires breaking the symmetry of the reflection coefficient with respect to $x$ or, equivalently, with respect to $k_x$. In other words, the reflection coefficient~\eqref{eq:reflection_coefficient} must be a non-even function of $k_x$, i.e.,
\begin{equation}\label{eq:R_kx_-kx}
R(k_x) \neq R(-k_x).
\end{equation}
Inspecting~\eqref{eq:reflection_coefficient} reveals that this condition requires
\begin{equation}\label{eq:normal_NR_relations}
\chi_\text{ee}^{xz} \neq \chi_\text{ee}^{zx}\quad \text{or} \quad \chi_\text{em}^{zy} \neq -\chi_\text{me}^{yz},
\end{equation}
which correspond to the first and third relations in~\eqref{eq:scalar_NR_relations}, respectively. Thus, breaking reciprocity in reflection can be accomplished only via \emph{normal} susceptibilities (under the prevailing nongyrotropy assumption~\cite{lavigne2021magnetless}). It can be shown that the second relation in~\eqref{eq:scalar_NR_relations}, involving tangential susceptibilities, breaks reciprocity in the $z$-direction~\cite{achouri2019angular}, which would be useful for transmission-type nonreciprocity.

\subsection*{Metasurface Design}
\subsubsection*{Susceptibility Derivation}
The specular isolator metasurface may be designed in the following three steps: i)~define the fields related to the desired wave transformations; ii)~insert these fields into~\eqref{eq:av_diff} to determine the appropriate field differences and averages; iii)~insert these last quantities into~\eqref{eq:GSTC_simplified_kx}, and solve the resulting equations for the susceptibility components. According to the analysis performed in~\secref{sec:required_susceptibility}, the susceptibility components obtained by this procedure should automatically respect the condition~\eqref{eq:normal_NR_relations}.

The field definitions in i) correspond here to the two field transformations represented in~\figref{fig:concept}. The first field transformation is the specular reflection of the wave incident in the $+z$-direction at the operation angle $\theta_0$ [\figref{fig:concept}~(a)]. The related fields are
\begin{subequations}\label{eq:transformation1}
\begin{equation}
\mathbf{E}_\text{1,i} =\cos \theta_0 e^{-j k_x x} \hat{x} - \sin \theta_0 e^{-j k_x x} \hat{z}, \quad \mathbf{H}_\text{1,i} =(e^{-j k_x x}/\eta )\hat{y},
\end{equation}
\begin{equation}
\mathbf{E}_\text{1,r} = -\cos \theta_0 e^{j \phi} e^{-j k_x x} \hat{x} - \sin \theta_0  e^{j \phi} e^{-j k_x x} \hat{z}, \quad \mathbf{H}_\text{1,r} =  (e^{j \phi} e^{-j k_x x}/\eta) \hat{y} ,
\end{equation}
\end{subequations}
where $\phi$ is the reflection phase induced by the metasurface. The second transformation is the absorption of the wave incident at the angle $-\theta_0$ [\figref{fig:concept}~(b)]. The related fields are
\begin{subequations}\label{eq:transformation2}
\begin{equation}
\mathbf{E}_\text{2,i} = \cos \theta_0 e^{-j k_x x} \hat{x} + \sin \theta_0 e^{-j k_x x} \hat{z},\quad \mathbf{H}_\text{2,i} = (e^{-j k_x x}/\eta) \hat{y},
\end{equation}
\begin{equation}
\mathbf{E}_\text{2,r} = 0, \quad \mathbf{H}_\text{2,r} = 0.
\end{equation}
\end{subequations}
%


Successively substituting both~\eqref{eq:transformation1} and~\eqref{eq:transformation2} into~\eqref{eq:av_diff}, according to ii), and inserting the resulting expressions into~\eqref{eq:GSTC_simplified_kx}, according to iii), leads a system of $2 \times 2 = 4$ scalar equations with 9 unknowns. This is an underdetermined system with an infinite number of possible sets of susceptibilities. Since the operation of the metasurface has been completely determined at $\theta_0$, these sets correspond to different responses at other (unspecified) angles of incidence, and represent therefore degrees of freedom, which may be generally leveraged in the design of the metaparticle. Among these degrees of freedom, the parameters $\chi_\text{em}^{xy}$ and $\chi_\text{me}^{yx}$ correspond to structural asymmetry along the $z$-direction~\cite{achouri2019angular,lavigne2021generalized},  which would imply considerable complexity in the metaparticle design. Therefore, we heuristically set these parameters to zero ($\chi_\text{em}^{xy} = \chi_\text{me}^{yx} = 0$). This reduces the number of unknowns to 7, which we shall maintain as degrees of freedom at this point. The resulting system of equations leads to the 2 explicit susceptibility solutions
\begin{subequations}\label{eq:susceptibility_solution}
\begin{equation}\label{eq:susceptibility_solution_a}
  \chi_\text{ee}^{xx} = \frac{-2 j (1+ e^{j \phi}) \sec \theta_0}{k},
\end{equation}
\begin{equation}\label{eq:susceptibility_solution_b}
  \chi_\text{ee}^{xz} = \frac{2 j e^{j \phi}\csc \theta_0}{k},
\end{equation}
and to the 2 constraint relations
\begin{equation}\label{eq:susceptibility_solution_c}
  \chi_\text{mm}^{yy} = -\frac{2j \cos \theta_0 + k \chi_\text{ee}^{zz} \sin^2 \theta_0 + e^{j \phi} k \sin \theta_0 (\chi_\text{ee}^{zx} \cos \theta_0 + \chi_\text{ee}^{zz} \sin \theta_0)}{(1+e^{j \phi})k},
\end{equation}
\begin{equation}\label{eq:susceptibility_solution_d}
  \chi_\text{em}^{zy} =- \frac{k \chi_\text{me}^{yz}+k \chi_\text{ee}^{zx} \cos \theta_0 + e^{j \phi} (k \chi_\text{me}^{yz} + 2j \cot \theta_0)}{(1+e^{j \phi})k},
\end{equation}
\end{subequations}
between the remaining 5 susceptibilities.

\subsubsection*{Transistor-loaded Metaparticle}

\begin{figure}[h!]
  \centering
  \includegraphics[width=\linewidth]{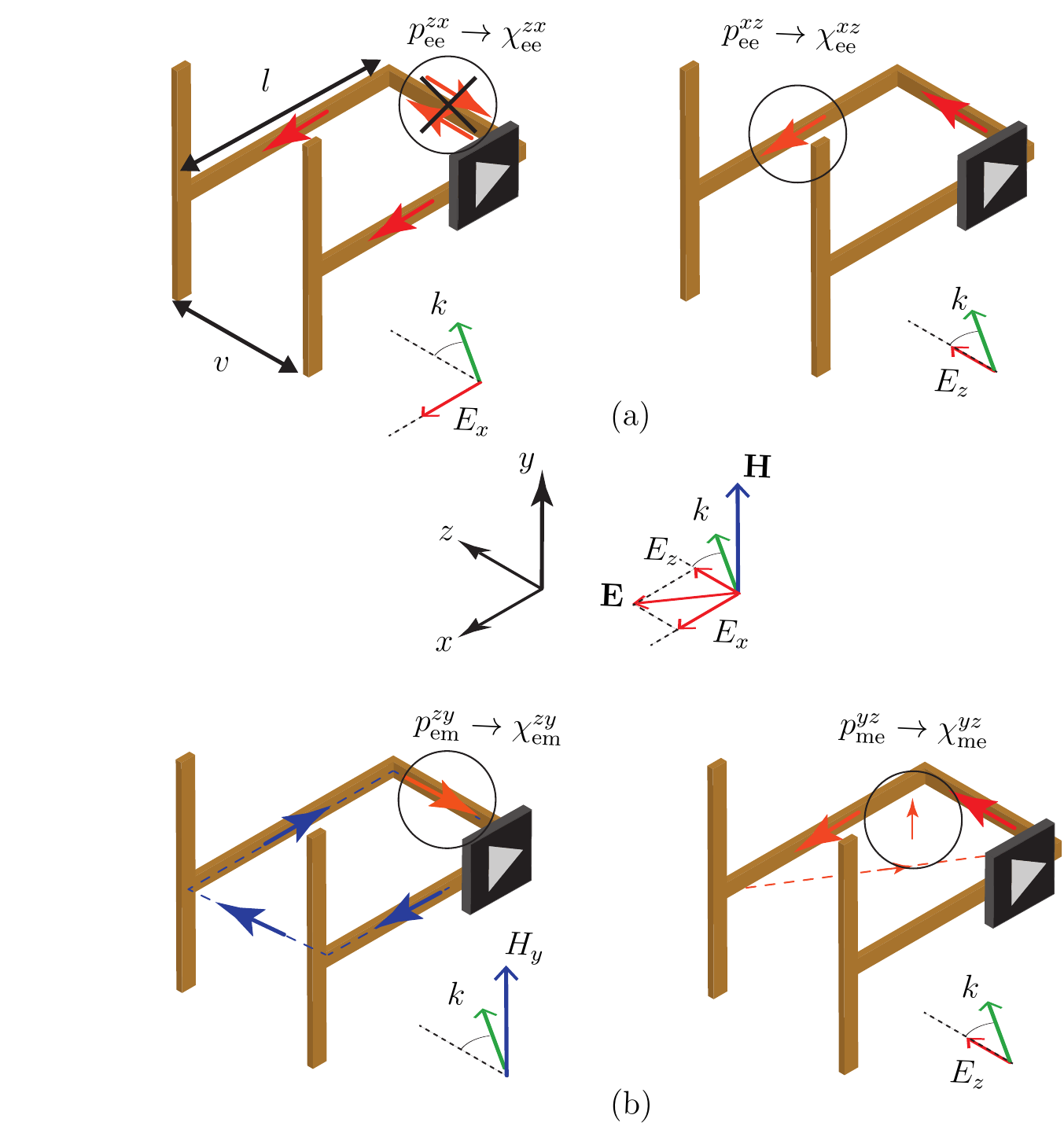}
  \caption{Operation principle of the transistor-loaded metaparticle, with orientation corresponding to~\figref{fig:concept}. (a)~Excitation $E_x$~(left) and $E_z$~(right), showing that $\chi_\text{ee}^{zx} \neq \chi_\text{ee}^{xz}$. (b)~Excitation from $H_y$~(left) and $E_z$~(right), showing that $\chi_\text{em}^{zy} \neq -\chi_\text{me}^{yz}$. The drawing is not to scale: in reality, $v \ll l$. The notation $p_\text{em}^{zy}$ represents the $z$ component of the electric dipole response due to the $y$ component of the magnetic field excitation, and so on. }\label{fig:metaparticle}
\end{figure}

The metaparticle structure satisfying the condition~\eqref{eq:normal_NR_relations} (nonreciprocity along the $x$-direction for p-polarization) and the relations~\eqref{eq:susceptibility_solution} (reflection and absorption at $\pm \theta_0$) may be devised as follows. Let us start by enforcing the first nonreciprocity condition in~\eqref{eq:normal_NR_relations}, namely $\chi_\text{ee}^{xz} \neq \chi_\text{ee}^{zx}$. This condition implies the existence of nonreciprocally related electric dipole responses along $x$ and $z$, which immediately suggests an L-shape conducting structure loaded by a transistor, operating as a unilateral element (e.g., common-source configuration in the case of a FET), in the $xz$-plane; this configuration is incidentally consistent with~\eqref{eq:susceptibility_solution_a} and~\eqref{eq:susceptibility_solution_b}. Such a structure implies in particular a $\chi_\text{ee}^{zz}$ response, which generally implies in turn a $\chi_\text{mm}^{yy}$ response according to~\eqref{eq:susceptibility_solution_c}. The latter corresponds to a $y$-directed magnetic dipole moment, which prompts us to close the L-shape into a loop in the $xz$-plane. We shall leave the loop open, as is customarily done for compactness in ring resonators, and we shall terminate the opened ends of the resulting U-shaped loop by T-shaped strips to reduce the size of the metaparticle. All these considerations lead to the metaparticle structure represented in \figref{fig:metaparticle}, which is composed of conducting strips in the three directions of space, with the spacing between the two $xy$-plane metallization planes being much smaller than the wavelength ($v \ll l < \lambda$, figure not to scale). We shall next analyze this metaparticle in details to verify that it indeed satisfies all the required conditions and to fully characterize it. Figure~\ref{fig:metaparticle} decomposes the excitations (incident fields) and responses (dipole moments) in order to determine how the metaparticle realizes the sought after nonreciprocal susceptibility components, although all of the excitations and  responses naturally occur simultaneously. Using this approach, we shall next examine the polarizability responses of the isolated metaparticle, which are directly related to the susceptibilities of the metasurface formed by its periodic repetition~\cite{achouri2020electromagnetic}.

Figure~\ref{fig:metaparticle}~(a) depicts the operation principle of the metaparticle realizing a response of the type $\chi_\text{ee}^{xz} \neq \chi_\text{ee}^{zx}$. On the left, the $x$-directed incident electric field induces in-phase ($v \ll \lambda$) currents in the two $x$-directed strips. When they reach the $z$-directed strip, these currents cancel out, which implies that the electric response along $z$ to the electric excitation along $x$ is zero ($p_\text{ee}^{zx} = 0 \rightarrow \chi_\text{ee}^{zx} =0$). On the right, the $z$-directed incident electric field induces a current in the $z$-directed strip. This induces a current only in one of the $x$-directed strips, given the orientation of the transistor, which implies that the electric response along $x$ to the electric field along $z$ is nonzero ($p_\text{ee}^{xz}\neq 0 \rightarrow \chi_\text{ee}^{xz} \neq 0$). Thus, the metaparticle corresponds to a specific set of solutions of~\eqref{eq:susceptibility_solution_c} and~\eqref{eq:susceptibility_solution_d} that is characterized by $\chi_\text{ee}^{zx} =0$, which simplifies these constraint equations to
\begin{subequations}\label{eq:susceptibility_sol_reduced}
  \begin{equation}\label{eq:susceptibility_sol_reduced_a}
  \chi_\text{mm}^{yy} = -\frac{2j \cos \theta_0 +(1+e^{j \phi}) k \chi_\text{ee}^{zz} \sin^2 \theta_0 }{(1+e^{j \phi})k},
\end{equation}
and
\begin{equation}\label{eq:susceptibility_sol_reduced_b}
  \chi_\text{em}^{zy} =-\chi_\text{me}^{yz} - \frac{2j e^{j \phi} \cot \theta_0 }{(1+e^{j \phi})k}.
\end{equation}
\end{subequations}

Equation~\eqref{eq:susceptibility_sol_reduced_b} reveals that the metaparticle must also satisfy the second nonreciprocity condition in~\eqref{eq:normal_NR_relations}. Let us see whether this is indeed the case with the help of~\figref{fig:metaparticle}~(b). On the left, the $y$-directed incident magnetic field induces a current in the metaparticle loop. The current flowing in the $z$-directed strip implies an electric response along $z$ due to the magnetic excitation along $y$ ($p_\text{em}^{zy} \rightarrow \chi_\text{em}^{zy}$). On the right, the $z$-directed incident electric field induces a current in $z$-directed strip. This can induce a current only in one of the two $x$-directed strips given the to the orientation of the transistor, which produces only a weak magnetic loop along $y$ ($p_\text{me}^{yz} \rightarrow \chi_\text{me}^{yz}$). This implies that $\chi_\text{em}^{zy} \neq -\chi_\text{me}^{yz}$, which is consistent with the requirement of~\eqref{eq:susceptibility_sol_reduced_b}.


We have thus found that the metasurface constituted of the heuristic metaparticle shown in~\figref{fig:metaparticle} breaks reciprocity in two distinct fashions, through $\chi_\text{ee}^{zx} \neq \chi_\text{ee}^{xz}$ and  $\chi_\text{me}^{yz} \neq \chi_\text{em}^{zy}$. These two types of nonreciprocity represent, both in isolation and in combination, novel metasurface nonreciprocal responses. Moreover, these responses, involving normal susceptibility components, were deemed particularly difficult to realize in practice~\cite{achouri2019angular}. The asymmetry of the electric susceptibility tensor, $\overline{\overline{\chi}}_\text{ee} \neq \overline{\overline{\chi}}_\text{ee}^\text{T}$, also appears in magnetized plasmas, but conjunctly with gyrotropy, while the non-antisymmetry between the magneto-electric susceptibility tensors, $\overline{\overline{\chi}}_\text{em} \neq - \overline{\overline{\chi}}_\text{me}^\text{T}$, also appears in the transmissive nonreciprocal metasurface in~\cite{taravati2017nonreciprocal}, but in terms of tangential nonreciprocal components.



\begin{figure}[h!]
  \centering
  \includegraphics[width=\linewidth]{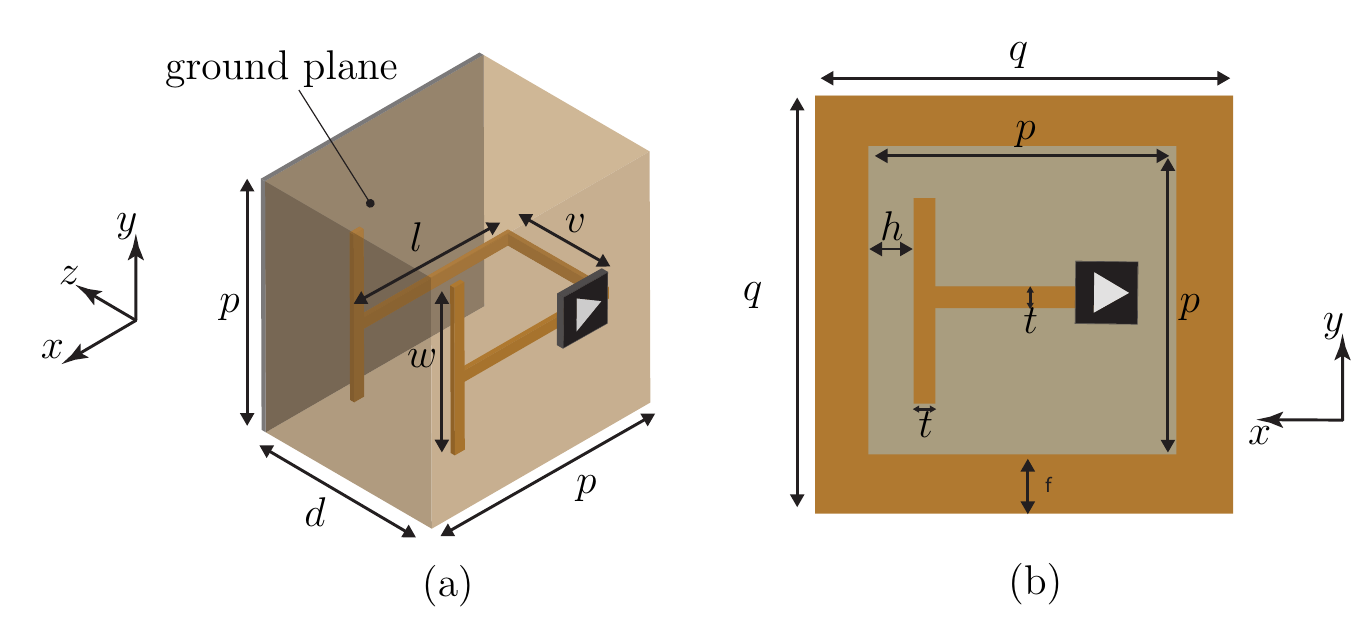}
  \caption{Unit cell corresponding to the metaparticle in~\figref{fig:metaparticle}, within a substrate of relative permittivity $\epsilon_\text{r} = 4.5$, backed by a ground plane and with a front conducting frame. (a) Perspective view (without the front frame, for visibility). (b) Top view.}\label{fig:unit_cell}
\end{figure}

Figure~\ref{fig:unit_cell} shows the metasurface unit cell of our experimental prototype. This unit cell corresponds to the metaparticle in~\figref{fig:metaparticle}, except for the additional supporting substrate, backing ground plane and conducting front frame, where the ground plane ensures impenetrability of the structure for all angles of incidence (extra specification) and the front frame isolates the unit cells from each other (hence ensuring direct correspondence between the polarizabilities and the susceptibilities). The parameters of the unit cell were optimized to satisfy~\eqref{eq:susceptibility_solution_a} and~\eqref{eq:susceptibility_solution_b} and one of the possible solutions of~\eqref{eq:susceptibility_sol_reduced}.


%

\subsection*{Simulation and Experiment}

\begin{figure}[h!]
  \centering
  \includegraphics[width=\linewidth]{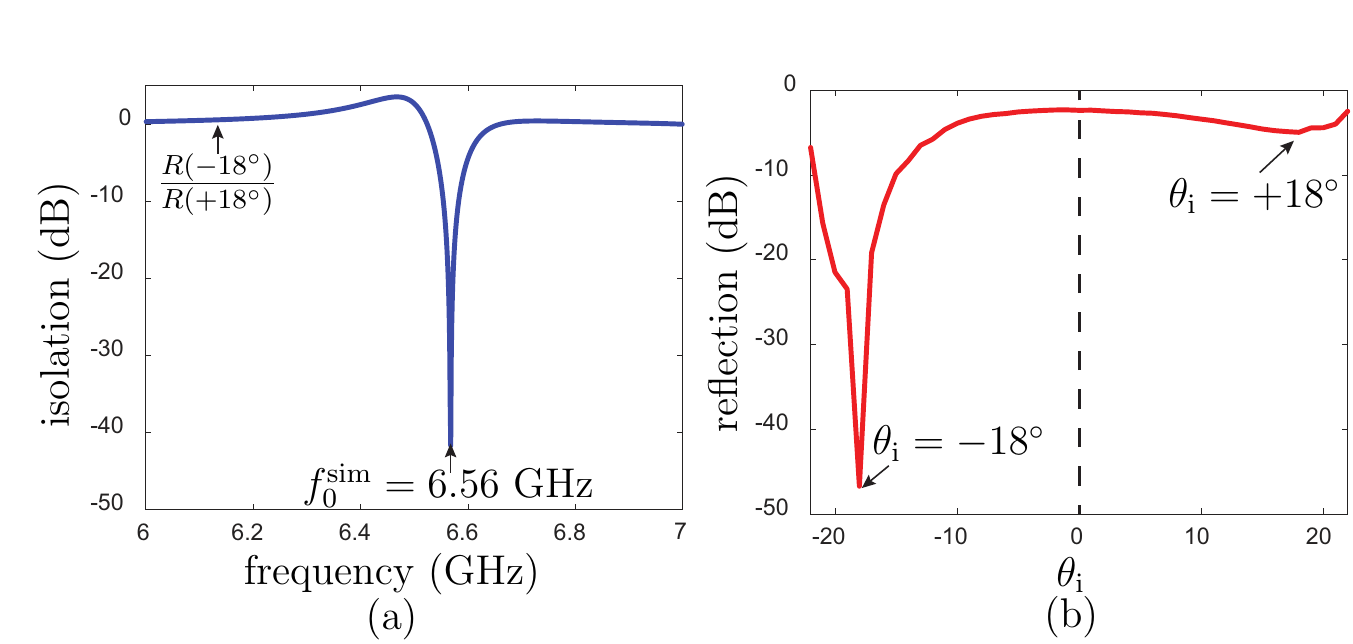}
  \caption{Full-wave simulation of the specular isolator with the unit cell in~\figref{fig:unit_cell} and parameters $p = 22.5$~mm, $w = 14$~mm, $l = 20$~mm, $v = 1.6$~mm, $d = 3.2$~mm, $q = 32.5$~mm, $h=1$~mm and $t = 2$~mm.  (a)~Isolation versus frequency for incidence at $\theta_\text{i} =  \pm 18^\circ$~(design angle of isolation). (b)~Specular reflection coefficient versus incidence angle ($\theta_\text{i}$) at the operating frequency $f_{0}^\text{sim}=6.58$~GHz.}\label{fig:sim_plot}
\end{figure}

The transistor-loaded unit cell in~\figref{fig:unit_cell} was simulated with periodic boundary conditions using a full-wave electromagnetic simulator (CST Microwave Studio) and the unilateral transistor circuit was modelled as an ideal isolator with a phase shifter. An FR4 slab with $\epsilon_r =4.5$ was used as the substrate and the geometrical parameters of the metasurface were optimized to realize the specular isolation operation. The metasurface was designed to provide specular isolation between the angles $\pm 18^\circ$ at the frequency of $f_{0}^\text{sim} = 6.56$~GHz for p-polarization. Figure~\ref{fig:sim_plot} presents the simulation results. Figure~\ref{fig:sim_plot}(a) shows the isolation response versus frequency, with the isolation $R(-18^\circ)/R(+18^\circ)$ (see \figref{fig:concept}) reaching $41.75$~dB at $f_{0}^\text{sim}$. Figure~\ref{fig:sim_plot}(b) shows the angular response of the reflection coefficient at the operating frequency of $f_{0}^\text{sim}$, whose strong asymmetry with respect to broadside ($\theta_\text{i} =0$) is the expected signature of the device. 

\begin{figure}[h!]
  \centering
\includegraphics[width=\linewidth]{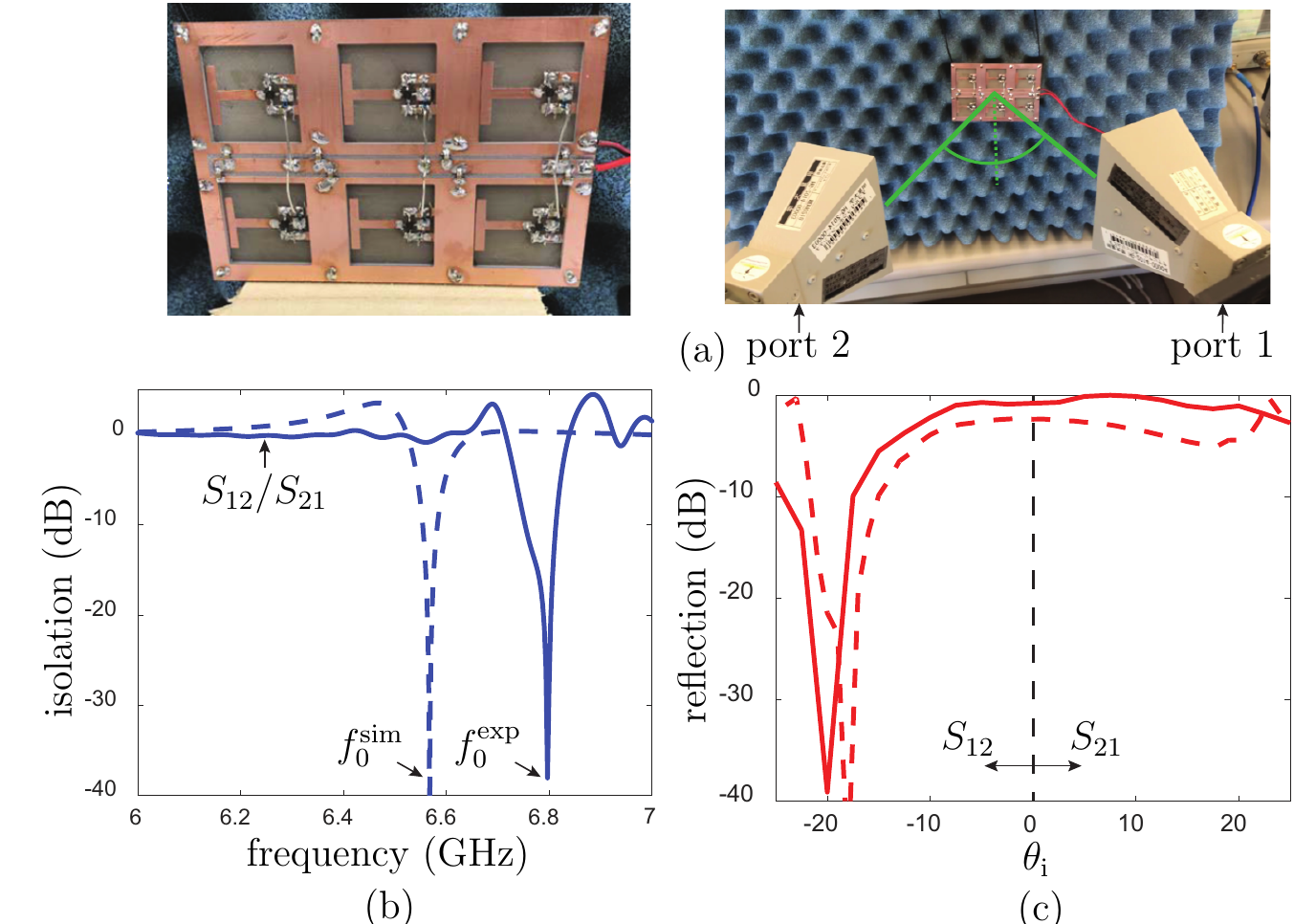}
  \caption{Experiment. (a)~Fabricated prototype (left) and experimental setup (right). (b)~Isolation versus frequency for incidence at $\theta_\text{i} = \pm 20^\circ$ (maximal isolation angle). (c)~Specular reflection coefficient versus incidence angle ($\theta_\text{i}$) the operating frequency of $f_{0}^\text{exp} = 6.797$~GHz. }\label{fig:exp_plot}
\end{figure}

Figure~\ref{fig:exp_plot} presents the experimental results. Figure~\ref{fig:exp_plot}(a) shows the prototype, composed of $2 \times 3$ unit cells. It includes two FR4 substrates of thickness $1.6$~mm glued together. The device was measured by a bistatic measurement system with two horn antennas symmetrically aiming (under the same angle with respect to the normal of the metasurface) at the metasurface. The reflection coefficient was measured for angles sweeping the sector extending $-22^\circ$ to $22^\circ$. Figure~\ref{fig:exp_plot}(b) shows the measured isolation ($|S_{12}|/|S_{21}|$) versus frequency for the incidence angle of $\theta_\text{i}= \pm 20^\circ$. An isolation of more than $38$~dB is observed at the frequency of $f_{0}^\text{exp} = 6.797$~GHz, whose discrepancy ($0.217$~GHz, i.e., $3.3\%$) may be explained by the small gap between the two substrates that was not taken into account in the simulation and by the difference between the actual transistor circuit response and the ideal isolator model used in simulation. Figure~\ref{fig:exp_plot}(c) shows the measured angular reflection coefficient at the operating frequency of $f_{0}^\text{exp}$. Here, the discrepancy translates into an angular difference ($2^\circ$).


%

\section*{Discussion}\label{sec:conclusion}

We reported a metasurface magnetless specular isolator. We derived, under the assumption of nongyrotropy, the corresponding bianisotropic susceptibility tensors, which include unusual, normal components, and which represent a novel type of nonreciprocity. We designed a metaparticle realizing these susceptibility tensors under the form of a U-shaped conducting structure loaded by a transistor, and demonstrated the overall metasurface by full-wave simulation and experimental results in the microwave regime.

Potential applications of this device include nonreciprocal reflectors, nonreciprocal waveguide walls, nonreciprocal quantum state mediation, advanced analog processing, as well as more sophisticated nonreciprocal electromagnetic wave transformations.

\bibliographystyle{IEEEtran}
\bibliography{LIB}

\begin{thebibliography}{10}
\providecommand{\url}[1]{#1}
\csname url@samestyle\endcsname
\providecommand{\newblock}{\relax}
\providecommand{\bibinfo}[2]{#2}
\providecommand{\BIBentrySTDinterwordspacing}{\spaceskip=0pt\relax}
\providecommand{\BIBentryALTinterwordstretchfactor}{4}
\providecommand{\BIBentryALTinterwordspacing}{\spaceskip=\fontdimen2\font plus
\BIBentryALTinterwordstretchfactor\fontdimen3\font minus
  \fontdimen4\font\relax}
\providecommand{\BIBforeignlanguage}[2]{{%
\expandafter\ifx\csname l@#1\endcsname\relax
\typeout{** WARNING: IEEEtran.bst: No hyphenation pattern has been}%
\typeout{** loaded for the language `#1'. Using the pattern for}%
\typeout{** the default language instead.}%
\else
\language=\csname l@#1\endcsname
\fi
#2}}
\providecommand{\BIBdecl}{\relax}
\BIBdecl

\bibitem{lax1962microwave}
B.~Lax and K.~J. Button, ``Microwave ferrites and ferrimagnetics,'' 1962.

\bibitem{villaverde1978terbium}
A.~B. Villaverde, D.~Donatti, and D.~Bozinis, ``Terbium gallium garnet {V}erdet
  constant measurements with pulsed magnetic field,'' \emph{J. Phys. C: Solid
  State Phys.}, vol.~11, no.~12, p. L495, 1978.

\bibitem{reiskarimian2021review}
N.~Reiskarimian, ``A review of nonmagnetic nonreciprocal electronic devices:
  Recent advances in nonmagnetic nonreciprocal components,'' \emph{IEEE
  Solid-State Circuits Magazine}, vol.~13, no.~4, pp. 112--121, 2021.

\bibitem{shi2015limitations}
Y.~Shi, Z.~Yu, and S.~Fan, ``Limitations of nonlinear optical isolators due to
  dynamic reciprocity,'' \emph{Nat. Photonics}, vol.~9, no.~6, pp. 388--392,
  2015.

\bibitem{sounas2018broadband}
D.~L. Sounas, J.~Soric, and A.~Al{\`u}, ``Broadband passive isolators based on
  coupled nonlinear resonances,'' \emph{Nat. Electron.}, vol.~1, no.~2, pp.
  113--119, 2018.

\bibitem{caloz2018electromagnetic}
C.~Caloz, A.~Al{\`u}, S.~Tretyakov, D.~Sounas, K.~Achouri, and Z.-L.
  Deck-L{\'e}ger, ``Electromagnetic nonreciprocity,'' \emph{Phys. Rev. Appl.},
  vol.~10, no.~4, p. 047001, 2018.

\bibitem{hadad2016breaking}
Y.~Hadad, J.~C. Soric, and A.~Al{\`u}, ``Breaking temporal symmetries for
  emission and absorption,'' \emph{Proc. Natl. Acad. Sci. U.S.A.}, vol. 113,
  no.~13, pp. 3471--3475, 2016.

\bibitem{shi2017optical}
Y.~Shi, S.~Han, and S.~Fan, ``Optical circulation and isolation based on
  indirect photonic transitions of guided resonance modes,'' \emph{ACS
  Photonics}, vol.~4, no.~7, pp. 1639--1645, 2017.

\bibitem{sounas2013giant}
D.~L. Sounas, C.~Caloz, and A.~Al{\`u}, ``Giant non-reciprocity at the
  subwavelength scale using angular momentum-biased metamaterials,'' \emph{Nat.
  Commun.}, vol.~4, no.~1, pp. 1--7, 2013.

\bibitem{sounas2018angular}
D.~L. Sounas, N.~A. Estep, A.~Kord, and A.~Al{\`u}, ``Angular-momentum biased
  circulators and their power consumption,'' \emph{IEEE Antennas Wirel. Propag.
  Lett.}, vol.~17, no.~11, pp. 1963--1967, 2018.

\bibitem{wang2018time}
Y.~E. Wang, ``On time-modulation-enabled nonreciprocity,'' \emph{IEEE Antennas
  Wirel. Propag. Lett.}, vol.~17, no.~11, pp. 1973--1977, 2018.

\bibitem{shaltout2015time}
A.~Shaltout, A.~Kildishev, and V.~Shalaev, ``Time-varying metasurfaces and
  lorentz non-reciprocity,'' \emph{Opt. Mater. Exp.}, vol.~5, no.~11, pp.
  2459--2467, 2015.

\bibitem{taravati2020full}
S.~Taravati and G.~V. Eleftheriades, ``Full-duplex nonreciprocal beam steering
  by time-modulated phase-gradient metasurfaces,'' \emph{Physical Review
  Applied}, vol.~14, no.~1, p. 014027, 2020.

\bibitem{popa2007architecture}
B.-I. Popa and S.~A. Cummer, ``An architecture for active metamaterial
  particles and experimental validation at rf,'' \emph{Microw. Opt. Technol.
  Lett}, vol.~49, no.~10, pp. 2574--2577, 2007.

\bibitem{yuan2009zero}
Y.~Yuan, B.-I. Popa, and S.~A. Cummer, ``Zero loss magnetic metamaterials using
  powered active unit cells,'' \emph{Opt. Express}, vol.~17, no.~18, pp.
  16\,135--16\,143, 2009.

\bibitem{popa2012nonreciprocal}
B.-I. Popa and S.~A. Cummer, ``Nonreciprocal active metamaterials,''
  \emph{Phys. Rev. B}, vol.~85, no.~20, p. 205101, 2012.

\bibitem{kodera2011artificial}
T.~Kodera, D.~L. Sounas, and C.~Caloz, ``Artificial {F}araday rotation using a
  ring metamaterial structure without static magnetic field,'' \emph{Appl.
  Phys. Lett}, vol.~99, no.~3, p. 031114, 2011.

\bibitem{wang2012gyrotropic}
Z.~Wang, Z.~Wang, J.~Wang, B.~Zhang, J.~Huangfu, J.~D. Joannopoulos,
  M.~Solja{\v{c}}i{\'c}, and L.~Ran, ``Gyrotropic response in the absence of a
  bias field,'' \emph{Proc. Natl. Acad. Sci. U.S.A}, vol. 109, no.~33, pp.
  13\,194--13\,197, 2012.

\bibitem{sounas2012electromagnetic}
D.~L. Sounas, T.~Kodera, and C.~Caloz, ``Electromagnetic modeling of a
  magnetless nonreciprocal gyrotropic metasurface,'' \emph{IEEE Trans. Antennas
  Propag.}, vol.~61, no.~1, pp. 221--231, 2012.

\bibitem{kodera2013magnetless}
T.~Kodera, D.~L. Sounas, and C.~Caloz, ``Magnetless nonreciprocal metamaterial
  (mnm) technology: application to microwave components,'' \emph{IEEE Trans.
  Microw. Theory Tech.}, vol.~61, no.~3, pp. 1030--1042, 2013.

\bibitem{ra2016magnet}
Y.~Ra'di and A.~Grbic, ``Magnet-free nonreciprocal bianisotropic
  metasurfaces,'' \emph{Phys. Rev. B}, vol.~94, no.~19, p. 195432, 2016.

\bibitem{taravati2017nonreciprocal}
S.~Taravati, B.~A. Khan, S.~Gupta, K.~Achouri, and C.~Caloz, ``Nonreciprocal
  nongyrotropic magnetless metasurface,'' \emph{IEEE Trans. Antennas Propag.},
  vol.~65, no.~7, pp. 3589--3597, 2017.

\bibitem{kodera2018unidirectional}
T.~Kodera and C.~Caloz, ``Unidirectional loop metamaterials (ulm) as magnetless
  artificial ferrimagnetic materials: principles and applications,'' \emph{IEEE
  Antennas Wirel. Propag. Lett.}, vol.~17, no.~11, pp. 1943--1947, 2018.

\bibitem{ra2020nonreciprocal}
Y.~Ra’di and A.~Al{\`u}, ``Nonreciprocal wavefront manipulation in
  synthetically moving metagratings,'' in \emph{Photonics}, vol.~7,
  no.~2.\hskip 1em plus 0.5em minus 0.4em\relax Multidisciplinary Digital
  Publishing Institute, 2020, p.~28.

\bibitem{lavigne2021magnetless}
G.~Lavigne and C.~Caloz, ``Magnetless reflective gyrotropic spatial isolator
  metasurface,'' \emph{New J. Phys.}, vol.~23, 2021.

\bibitem{taravati2021full}
S.~Taravati and G.~V. Eleftheriades, ``Full-duplex reflective beamsteering
  metasurface featuring magnetless nonreciprocal amplification,'' \emph{Nat.
  Commun.}, vol.~12, p. 4414, 2021.

\bibitem{li2022nonreciprocal}
Y.~B. Li, S.~Y. Wang, H.~P. Wang, H.~Li, J.~L. Shen, and T.~J. Cui,
  ``Nonreciprocal control of electromagnetic polarizations applying active
  metasurfaces,'' \emph{Adv. Opt. Mater.}, p. 2102154, 2022.

\bibitem{glybovski2016metasurfaces}
S.~B. Glybovski, S.~A. Tretyakov, P.~A. Belov, Y.~S. Kivshar, and C.~R.
  Simovski, ``Metasurfaces: From microwaves to visible,'' \emph{Phys. Rep.},
  vol. 634, pp. 1--72, 2016.

\bibitem{achouri2018design}
K.~Achouri and C.~Caloz, ``Design, concepts, and applications of
  electromagnetic metasurfaces,'' \emph{Nanophotonics}, vol.~7, no.~6, pp.
  1095--1116, 2018.

\bibitem{pfeiffer2014bianisotropic}
C.~Pfeiffer and A.~Grbic, ``Bianisotropic metasurfaces for optimal polarization
  control: Analysis and synthesis,'' \emph{Phys. Rev, Appl.}, vol.~2, no.~4, p.
  044011, 2014.

\bibitem{epstein2016arbitrary}
A.~Epstein and G.~V. Eleftheriades, ``Arbitrary power-conserving field
  transformations with passive lossless omega-type bianisotropic
  metasurfaces,'' \emph{IEEE Trans. Antennas Propag.}, vol.~64, no.~9, pp.
  3880--3895, 2016.

\bibitem{asadchy2016perfect}
V.~S. Asadchy, M.~Albooyeh, S.~N. Tcvetkova, A.~D{\'\i}az-Rubio, Y.~Ra'di, and
  S.~Tretyakov, ``Perfect control of reflection and refraction using spatially
  dispersive metasurfaces,'' \emph{Phys. Rev. B}, vol.~94, no.~7, p. 075142,
  2016.

\bibitem{lavigne2018susceptibility}
G.~Lavigne, K.~Achouri, V.~S. Asadchy, S.~A. Tretyakov, and C.~Caloz,
  ``Susceptibility derivation and experimental demonstration of refracting
  metasurfaces without spurious diffraction,'' \emph{IEEE Trans. Antennas
  Propag.}, vol.~66, no.~3, pp. 1321--1330, 2018.

\bibitem{chen2018theory}
M.~Chen, E.~Abdo-S{\'a}nchez, A.~Epstein, and G.~V. Eleftheriades, ``Theory,
  design, and experimental verification of a reflectionless bianisotropic
  huygens' metasurface for wide-angle refraction,'' \emph{Phys. Rev. B},
  vol.~97, no.~12, p. 125433, 2018.

\bibitem{asadchy2018bianisotropic}
V.~S. Asadchy, A.~D{\'\i}az-Rubio, and S.~A. Tretyakov, ``Bianisotropic
  metasurfaces: physics and applications,'' \emph{Nanophotonics}, vol.~7,
  no.~6, pp. 1069--1094, 2018.

\bibitem{achouri2015general}
K.~Achouri, M.~A. Salem, and C.~Caloz, ``General metasurface synthesis based on
  susceptibility tensors,'' \emph{IEEE Trans. Antennas Propag.}, vol.~63,
  no.~7, pp. 2977--2991, 2015.

\bibitem{epstein2016huygens}
A.~Epstein and G.~V. Eleftheriades, ``Huygens’ metasurfaces via the
  equivalence principle: design and applications,'' \emph{JOSA B}, vol.~33,
  no.~2, pp. A31--A50, 2016.

\bibitem{achouri2020electromagnetic}
K.~Achouri and C.~Caloz, \emph{Electromagnetic Metasurfaces: Theory and
  Applications}.\hskip 1em plus 0.5em minus 0.4em\relax Wiley-IEEE Press, 2020.

\bibitem{achouri2019angular}
K.~Achouri and O.~J. Martin, ``Angular scattering properties of metasurfaces,''
  \emph{IEEE Trans. Antennas Propag.}, vol.~68, no.~1, pp. 432--442, 2019.

\bibitem{abdolali2019parallel}
A.~Abdolali, A.~Momeni, H.~Rajabalipanah, and K.~Achouri, ``Parallel
  integro-differential equation solving via multi-channel reciprocal
  bianisotropic metasurface augmented by normal susceptibilities,'' \emph{New
  J. Phys.}, vol.~21, no.~11, p. 113048, 2019.

\bibitem{idemen2011discontinuities}
M.~M. Idemen, \emph{Discontinuities in the electromagnetic field}.\hskip 1em
  plus 0.5em minus 0.4em\relax John Wiley \& Sons, 2011, vol.~40.

\bibitem{kuester2003averaged}
E.~F. Kuester, M.~A. Mohamed, M.~Piket-May, and C.~L. Holloway, ``Averaged
  transition conditions for electromagnetic fields at a metafilm,'' \emph{IEEE
  Trans. Antennas Propag.}, vol.~51, no.~10, pp. 2641--2651, 2003.

\bibitem{lavigne2021generalized}
G.~Lavigne and C.~Caloz, ``Generalized brewster effect using bianisotropic
  metasurfaces,'' \emph{Opt. Express}, vol.~29, no.~7, pp. 11\,361--11\,370,
  2021.

\end{thebibliography}

\end{document}